\documentclass[letterpaper,twocolumn,10pt]{article}
\usepackage{graphicx}
\usepackage{subcaption}
\usepackage{usenix,epsfig,bytefield}
\usepackage{ifthen}
\usepackage{xcolor}
\usepackage{booktabs}
\usepackage{color}
\usepackage{colortbl}
\usepackage[T1]{fontenc}
\usepackage{todonotes}
\usepackage{authblk}

\usepackage{float}                           
\usepackage{bigdelim}

\usepackage[hyphens]{url}
\usepackage[colorlinks,citecolor=blue,breaklinks=true]{hyperref}%
\usepackage{svg}
\usepackage{lipsum}
\usepackage{amsmath}
\usepackage{amssymb}

\usepackage{soul}

\usepackage{datetime2}
\usepackage{tikz}
\usetikzlibrary{arrows.meta,decorations.pathreplacing,matrix,positioning,shapes.geometric,calc}
\newcommand{\showComments}{no}

\usepackage{multirow}
\usepackage[frozencache,cachedir=minted]{minted}
\usepackage{listings}
\definecolor{ZeroOSBlue}{HTML}{0000FF}
\definecolor{ZeroOSRed}{HTML}{FF0000}     
\definecolor{ZeroOSPurple}{HTML}{800080}  

\lstdefinestyle{CallStack}{
    basicstyle=\ttfamily,  
    tabsize=4,
    columns=fixed,               
    literate={CALLINTO}{$\hookrightarrow$}{1},
    keywords={_start, boot_kernel, build_musl_stack, __libc_start_main, __init_libc, __init_tls, __init_ssp, libc_start_main_stage2, __libc_start_init, libc_start_init, _init, main},
    keywordstyle=\color{ZeroOSBlue}\bfseries, 
    morekeywords={_start, boot_kernel, build_musl_stack, __libc_start_main, __init_libc, __init_tls, __init_ssp, libc_start_main_stage2, __libc_start_init, libc_start_init, _init, main}
}

\title{\textbf{ZeroOS: A Universal Modular Library OS for zkVMs}}

\author[1]{Guangxian Zou}
\author[1]{Isaac Zhang}
\author[1]{Ryan Zarick}
\author[1]{Kelvin Wong}
\author[1]{Thomas Kim}
\author[1]{Daniel~L.-K.~Wong}
\author[1]{Saeid Yazdinejad}
\author[2]{Dan Boneh}

\affil[1]{LayerZero Labs}
\affil[2]{Stanford University}
\date{}

\begin{document}

\widowpenalty10000
\clubpenalty10000

\maketitle

\ifthenelse{\equal{\showComments}{yes}}{
\SetWatermarkText{CONFIDENTIAL}
\SetWatermarkScale{.5}
\tikz[overlay,remember picture]{
\node at ($(current page.west)+(0.4,0)$) [rotate=90] {\Huge\textcolor{gray}{\DTMnow}};}}{}

\begin{abstract}
Zero-knowledge Virtual Machines (zkVMs) promise general-purpose verifiable computation through ISA-level compatibility with modern programs and toolchains.
However, compatibility extends further than just the ISA; modern programs often cannot run or even compile without an operating system and \texttt{libc}.
zkVMs attempt to address this by maintaining forks of language-specific runtimes and statically linking them into applications to create self-contained unikernels, but this ad-hoc approach leads to version hell and burdens verifiable applications (vApps) with an unnecessarily large trusted computing base.
We solve this problem with ZeroOS, a modular library operating system (libOS) for vApp unikernels; vApp developers can use off-the-shelf toolchains to compile and link only the exact subset of the Linux ABI their vApp needs.
Any zkVM team can easily leverage the ZeroOS ecosystem by writing a ZeroOS bootloader for their platform, resulting in a reduced maintainence burden and unifying the entire zkVM ecosystem with consolidated development and audit resources.
ZeroOS is free and open-sourced at \url{https://github.com/LayerZero-Labs/ZeroOS}.
\end{abstract}

\section{Introduction}
\label{sec:introduction}

Zero-knowledge Virtual Machines (zkVMs) cryptographically prove the correctness of execution of an application (called a \emph{guest}) against a given input with regard to the application semantics.
Guest proving on zkVM is commonly paired with commitment and verification of the proof on a byzantine-fault tolerant ledger to form a \emph{vApp}~\cite{zhang2025vapps}.

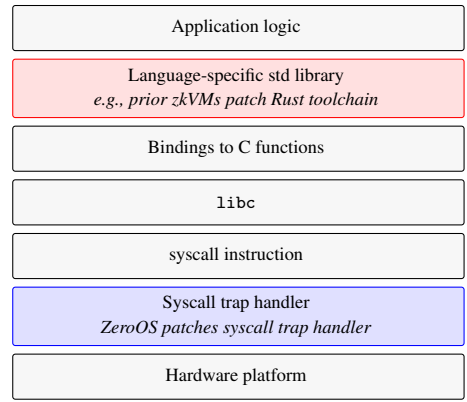
\begin{figure}[t]
    \centering
    \begin{tikzpicture}[
        stack/.style={
            matrix of nodes,
            nodes={
                draw,
                rounded corners=1pt,
                minimum width=6cm,
                minimum height=0.6cm,
                font=\scriptsize,
                align=center,
                fill=black!3
            },
            row sep=1mm
        },
        patch/.style={fill=ZeroOSRed!12, draw=ZeroOSRed},
        zeroos/.style={fill=ZeroOSBlue!12, draw=ZeroOSBlue}
    ]
        \matrix (stack) [stack] {
            Application logic \\
            |[patch]| {\shortstack{Language-specific std library\\\textit{e.g., prior zkVMs patch Rust toolchain}}} \\
            Bindings to C functions \\
            \texttt{libc} \\
            syscall instruction \\
            |[zeroos]| {\shortstack{Syscall trap handler\\\textit{ZeroOS patches syscall trap handler}}} \\
            Hardware platform \\
        };
    \end{tikzpicture}
    \caption{\textbf{ZeroOS versus status quo.} Prior zkVM toolchains patch each language's standard library (\textcolor{ZeroOSRed}{red}), whereas ZeroOS patches the syscall trap handler (\textcolor{ZeroOSBlue}{blue}), which is a generic and language-agnostic solution.}
    \label{fig:software-stack}
\end{figure}
While zkVMs promise to prove any application compiled for an ISA (e.g., RISC-V), reality falls short: ISA-level compatibility does not imply software compatibility.
Current zkVM execution environments are much closer to a bare-metal, no-OS environment that is incompatible with modern applications that rely on rich package ecosystems.

To bridge this gap, developers currently maintain forked, platform and language-specific toolchains to statically compile guests into self-contained \emph{unikernels}~\cite{10.1145/2490301.2451167}.
This approach requires many each zkVM team to maintain a separate toolchain for each supported language; as a result, vApp developers must accept the risk of bugs in an unnecessarily large \textit{trusted computing base}~\cite{nibaldi1979specification} of modified standard library code, and each zkVM team is burdened with costly audits for their platform-specific forked toolchains.

We solve this problem with our system ZeroOS, a universal modular library OS (libOS) for zkVM guest unikernels.
ZeroOS shifts the integration point from scattered language-specific forked runtimes to a language-agnostic syscall shim that implements a simplified subset of the Linux syscall ABI.
Developers can use off-the-shelf toolchains to link unmodified applications against ZeroOS, effectively eliminating ``version hell'' and unlocking universal compatibility between all applications and all zkVMs.
To illustrate the difference: custom toolchains directly compile \texttt{malloc} calls to jump to a custom \texttt{malloc} handler, whereas ZeroOS traps the syscall instruction and dispatches it to the \texttt{sys\_malloc} handler according to the ISA semantics (e.g., \texttt{ecall}, \texttt{SVC}, \texttt{syscall}).

Our architecture is a Pareto improvement over current monolithic or bare-metal approaches. vApp developers can build their application on top of POSIX~\cite{walli1995posix} interfaces, and existing applications can be recompiled to vApps without modification.
Each subsystem in ZeroOS (e.g., memory allocation, I/O) is modularized, allowing developers to choose the exact set of features they need to efficiently run their vApp.
Additionally, a single subsystem can have multiple implementations in different modules, so applications can choose between highly efficient barebones implementations and more feature-rich implementations by just swapping the module imports.
This ``pay-for-what-you-use'' model has two major benefits: (1) it gives developers full control over their exposure to software bugs in ZeroOS, and (2) it allows developers to minimize execution trace length by unplugging unused logic (e.g., I/O).

In practice, ZeroOS improves the security of \emph{\textbf{all}} vApps and \emph{\textbf{all}} zkVMs by consolidating ecosystem resources on a single shared codebase.
ZeroOS allows the industry to pool resources by standardizing a platform-agnostic OS layer: a single audit, security patch, or performance optimization to ZeroOS benefits all \emph{\textbf{all}} zkVMs and vApps simultaneously.
This replaces the fragmented landscape of ad-hoc patched toolchains with a unified, robust foundation for the entire growing ecosystem of over twenty zkVM projects~\cite{awesomezkvm}.

ZeroOS is open-source and available at \url{https://github.com/LayerZero-Labs/ZeroOS}.

\section{Background}

\subsection{ISA, operating system (OS), and standard library}
Modern software stacks are layered. At the bottom, the instruction set architecture (ISA) defines the semantics of a set of instructions (e.g., RISC-V) that make up the interface between hardware and software. Above that, an operating system (e.g., Linux) manages and exposes hardware resources to userspace applications through a system call (syscall) interface.
The system C library (\texttt{libc}) provides the fundamental userspace runtime and bindings to these syscalls. Language-specific standard libraries (e.g., Rust \texttt{std}) then build higher-level APIs for files, threads, and networking on top.

Importantly, there is a distinction between the POSIX-like syscall API exposed by libc to user applications and the OS-level ABI called by libc.
For example, \texttt{mmap} is a POSIX syscall that internally emits a syscall instruction (e.g., \texttt{ECALL}) to invoke the OS-level \texttt{sys\_mmap}.
However, \texttt{mmap} does not always invoke \texttt{sys\_mmap}--for efficiency, small allocations are often managed by libc entirely in userspace after initially provisioning a larger block of memory from the OS.

\begin{figure}
    \centering
    \includegraphics[width=1.0\linewidth]{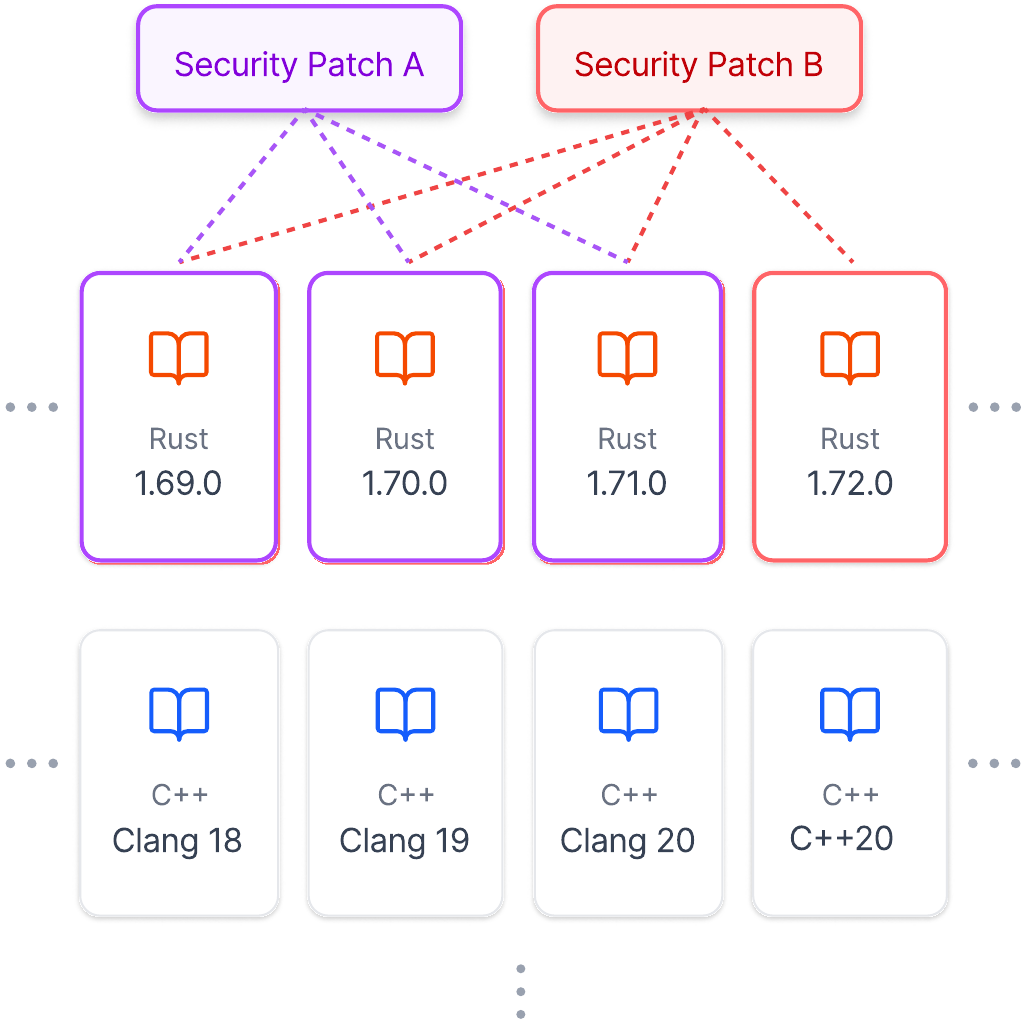}
    \caption{\textbf{Version hell.} Using a forked toolchain, each zkVM may need to maintain many different versions across multiple languages. Security patches also need to be rapidly backported to older versions to support customers who have not yet upgraded, creating a near-intractable long-term operational burden.}
    \label{fig:security-version-hell}
\end{figure}

\subsection{zkVM determinism and traces}

In this paper, we use the term \emph{guest} for all logic that runs atop the zkVM: this includes the \emph{application} (high level business logic), its language runtime, the C standard library, the ZeroOS bootloader, and the ZeroOS kernel. 

A \emph{prover} runs the guest in a zkVM, which executes the guest bytecode against some ISA specification and produces a succinct proof that this execution is correct.
During execution, the prover records a complete trace of guest execution steps, including register updates, memory accesses, and I/O.
The prover uses this trace as the witness to the zkVM's proof system, and a verifier checks the resulting proof without re-running the guest.
Therefore, the prover cost is tied to the length of the guest trace, and any non-deterministic influence from the host---such as time or randomness---must pass through an explicit interface and be modeled in the statement being proved.
As a result, current zkVM designs require guests to be fully deterministic, with a well-specified boundary between guest and host.

A challenge of developing zkVM guests (including ZeroOS) is that all guest logic must be fully \emph{constrained}--a malicious prover must not be capable of influencing the guest trace in a way that violates any guest-defined safety invariants.
This requirement affects ZeroOS in several subtle ways, precluding features with stateful side effects, and requiring inputs such as entropy and time to be deterministically constrained.

\subsection{Library OS and Unikernels}
A library operating system (libOS)~\cite{10.1145/224057.224076} implements operating-system functionality as a set of libraries linked into the application's address space, rather than as a separate monolithic kernel.
Unikernel~\cite{10.1145/2490301.2451167} systems such as Unikraft~\cite{kuenzer2021unikraft} statically link an application against a libOS into a standalone binary that runs directly on a hypervisor or bare metal.
Unikernels allow developers to select only the OS components needed by their application to improve predictability and efficiency.

Well-designed operating systems, including libOS and unikernels, are agnostic to guest application language and platform.
Linkage against the operating system interface is done after compilation from the high level language into machine code; guest applications written in any language that can be compiled to machine code can be linked against ZeroOS, which is a unique advantage of ZeroOS over forked toolchains.
While it is possible to make a platform-agnostic forked toolchain, in reality most forked toolchains contain platform-specific optimizations; ZeroOS is compiled alongside the application code using an off-the-shelf toolchain, making it inherently platform-agnostic.

\begin{figure*}[t]
    \centering
    \begin{subfigure}[b]{0.45\textwidth}
        \centering
        \includegraphics[width=\linewidth]{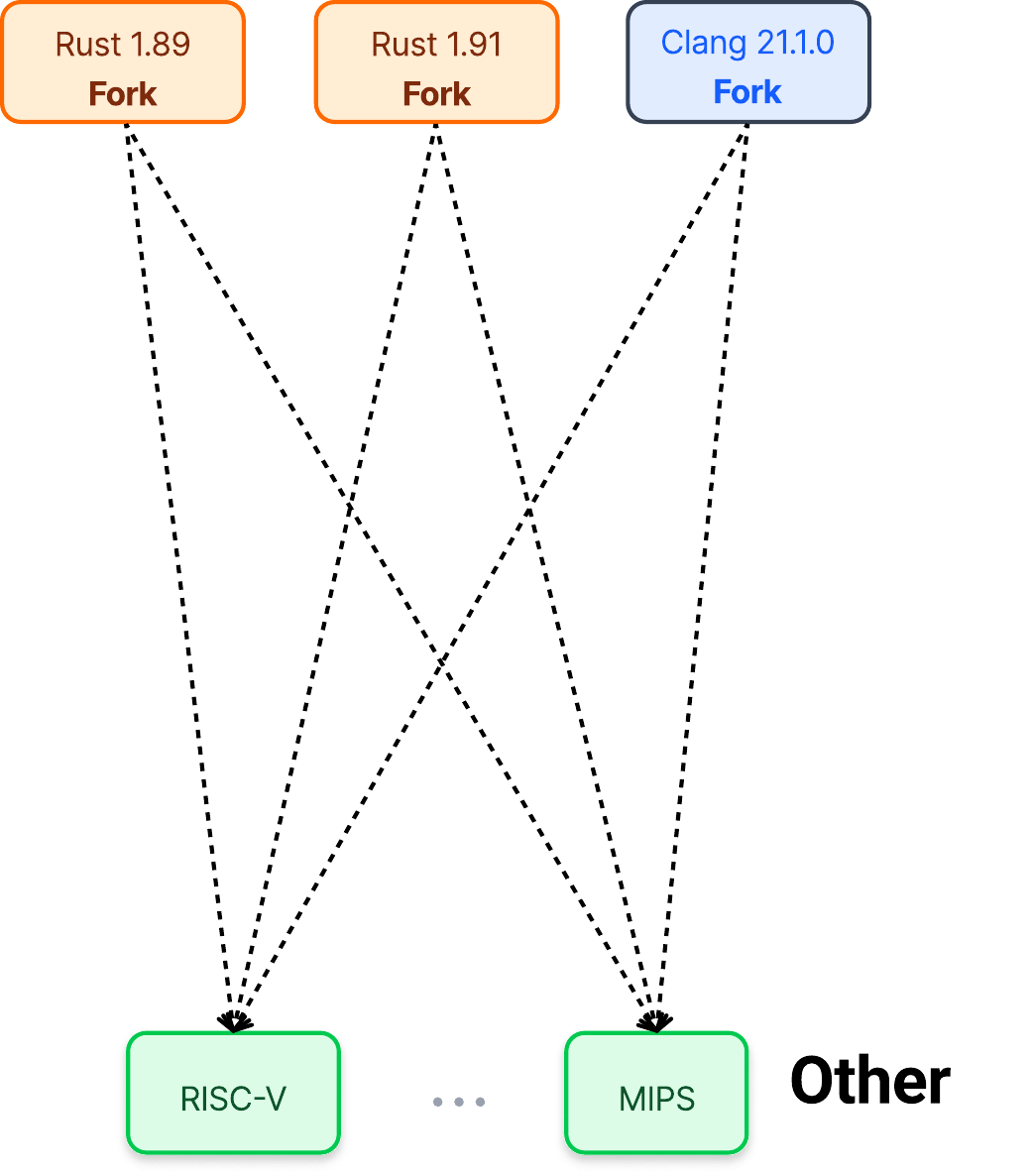}
        \caption{\textbf{Status quo:} many patched forks of language-specific toolchains and standard libraries compile directly to machine code.}
        \label{fig:other-architecture}
    \end{subfigure}
    \hfill
    \begin{subfigure}[b]{0.45\textwidth}
        \centering
        \includegraphics[width=\linewidth]{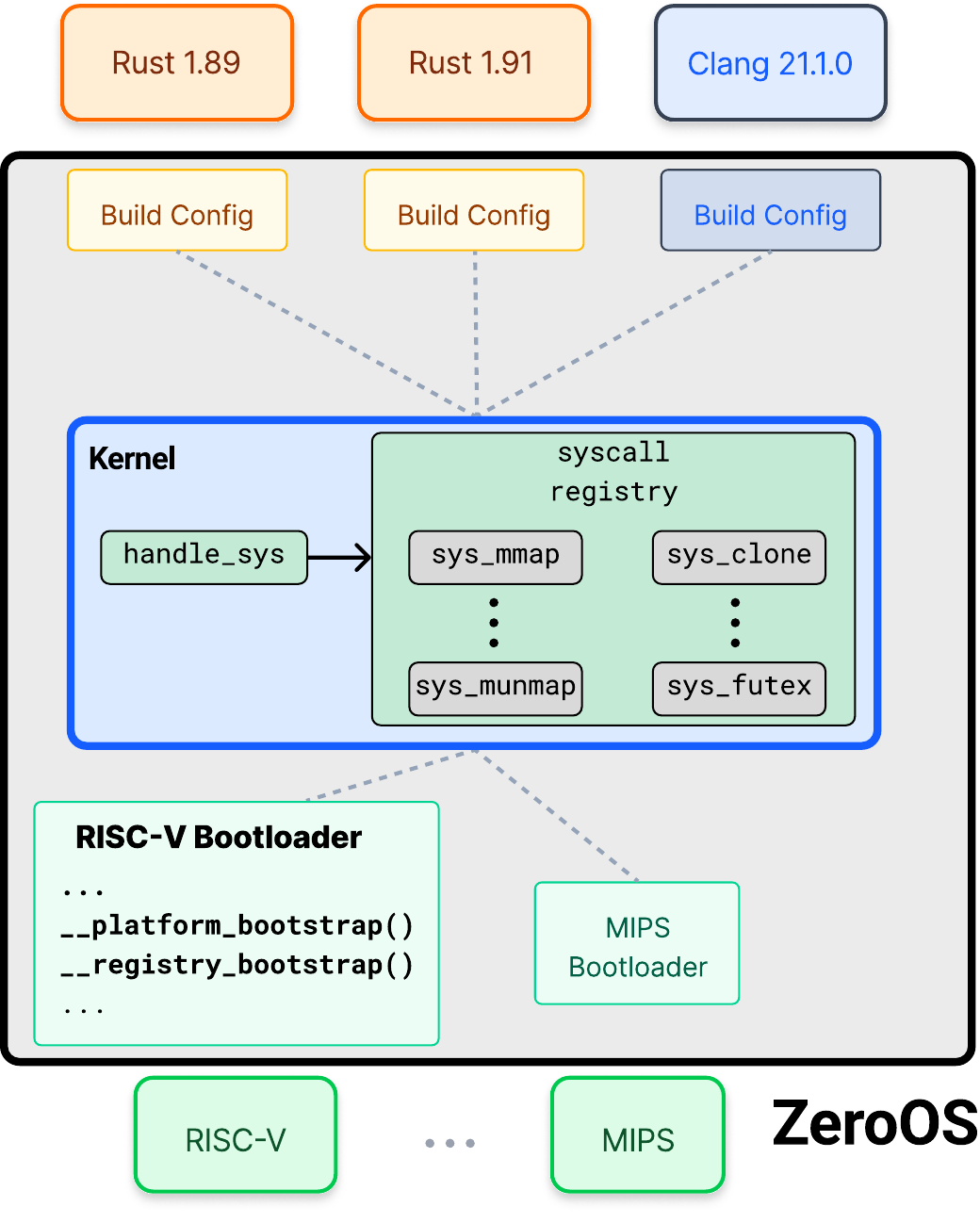}
        \caption{\textbf{ZeroOS:} modular kernel compatible with all toolchains and all platforms via build configuration and platform-specific bootloader respectively.}
        \label{fig:zero-os-architecture}
    \end{subfigure}
    \caption{ZeroOS minimizes platform-specific and language-specific code, improving security and simplifying maintenance.}
    \label{fig:architecture-comparison}
\end{figure*}

\subsection{Custom toolchains}
\label{sec:background-custom-toolchains}

Having set out the execution model and the libOS/unikernel patterns, we describe how zkVMs are typically engineered today.
Rather than handling normally emitted syscall instructions at the syscall shim layer, existing zkVM stacks modify language-specific toolchains and standard libraries to replace syscall instructions with a direct invocation of the syscall logic itself. We refer to these modifications collectively as \emph{custom toolchains}.

\vspace{1em}\noindent{\textbf{Ecosystem Fragmentation}}
The reliance on custom toolchains creates two fundamental problems for the zkVM ecosystem: operational fragmentation and security fragility. First, this approach creates \emph{language-specific development silos}. Each zkVM team is forced to maintain a separate fork of the toolchain and standard library for every language they support. This results in ``Version Hell'' (Figure~\ref{fig:security-version-hell}), where zkVM toolchains perpetually suffer from \emph{upstream drift}, lagging behind official releases. Developers are frequently blocked from using modern language features or security fixes because the zkVM-specific fork has not yet forward-ported the necessary patches from the upstream compiler.

\vspace{1em}\noindent{\textbf{Expanded Trusted Computing Base}}: Modifying the standard library fragments the security landscape. Every patch applied to a language runtime moves logic from the verified, community-audited upstream codebase into a custom, unaudited fork. A bug in a custom memory allocator patch or a modified thread shim can introduce vulnerabilities that are unique to that specific zkVM implementation. By relying on ad-hoc patches, the industry currently scatters its security resources across dozens of divergent forks.

The slow security response worsens the situation: When a security advisory is published for Rust, developers cannot use the fix until it is ported to the necessary patched fork(s), significantly widening the vulnerability window for vApps.
We believe the more secure and scalable approach is to \emph{minimize} the trusted computing base and push all modifications lower in the stack (into the kernel) where interfaces are simplest and most stable.

\section{Design}
\label{sec:design}

\begin{figure}[t]
    \centering
    \includegraphics[width=\linewidth]{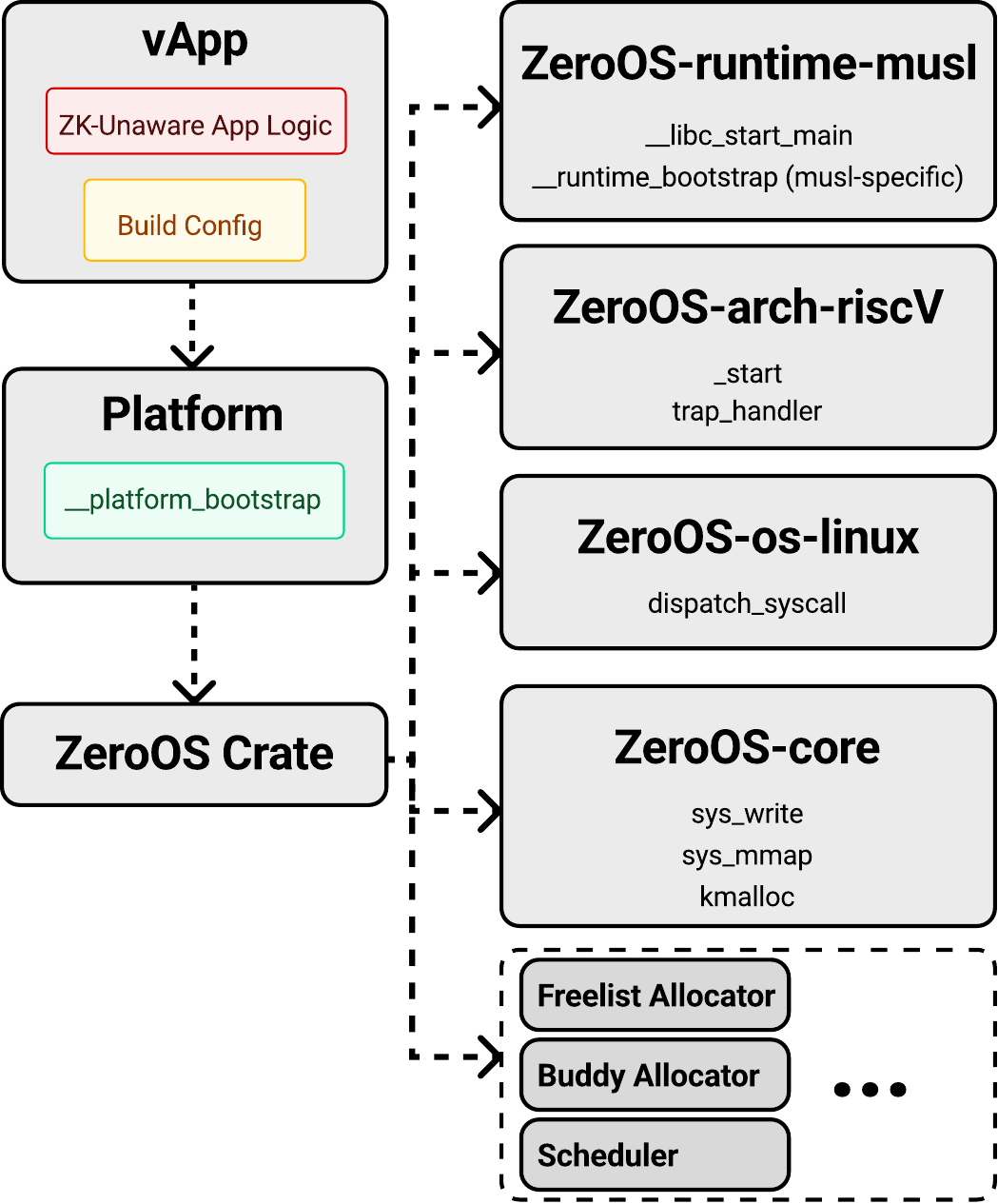}
    \caption{The ZeroOS package is a collection of packages, including libc (ZeroOS-runtime-musl), bootloader (ZeroOS-arch-riscV), syscall ABI (ZeroOS-os-linux), syscall wrappers (ZeroOS-core), and various implementations of OS primitives (e.g., freelist allocator) in separate modular sub-packages. Each zkVM platform must implement \texttt{\_\_platform\_bootstrap} which is called by bootloader \texttt{\_start}. vApps automatically inherit this entire stack when they import the platform (zkVM) package into their project.}
    \label{fig:zero-os-crates}
\end{figure}

ZeroOS is a library OS that enables compilation of traditional applications into verifiable applications (vApps)~\cite{zhang2025vapps} packaged as unikernels.
Unlike existing approaches which rely on custom language-specific toolchains, ZeroOS performs this transparently at the syscall shim layer \emph{without modifying libc}.
As a result, the ZeroOS kernel is language and platform agnostic, and can be compiled using an unmodified off-the-shelf toolchain (note that some toolchains may have to be rebuilt for the target platform (e.g., \texttt{riscv64ima-unknown-linux-musl}) without modification to the source code).
ZeroOS-specific changes to the application are limited to build scripts within each toolchain (e.g., \texttt{musl-toolchain.sh}) rather than the toolchain itself.
This design minimizes the trusted computing base and eliminates toolchain fragmentation; additionally, the majority of nontrivial logic is isolated to the highly modular kernel, which enables rapid response to security vulnerabilities.
Figure~\ref{fig:zero-os-crates} illustrates how developers using ZeroOS can link their zkVM-unaware application logic to ZeroOS to transform off-the-shelf applications into vApps.

Figure~\ref{fig:zero-os-architecture} illustrates the low-level architecture of ZeroOS with three main components: a small \emph{\textbf{build config}} for each toolchain to statically link the application against ZeroOS, a \emph{\textbf{kernel}} that defines the syscall trap handler and syscall handlers, and a \emph{\textbf{bootloader}} which performs all necessary architecture-specific bootstrapping steps.
This design contrasts with the currently ubiquitous approach of simply maintaining a separate zkVM-specific fork of each toolchain, which as we described in Section~\ref{sec:background-custom-toolchains}, is impractical due to the operational burden of maintaining many versions across many languages.

\subsection{Build Config}
The build config acts as the \emph{universal compatibility} bridge between the \emph{unmodified} language-specific toolchain (e.g., preprocessor, compiler, linker) and ZeroOS kernel.
ZeroOS allows developers to simply configure their existing toolchain to statically link their application against the ZeroOS bootloader and kernel to build a unikernel.
By isolating ZeroOS-specific changes to build scripts rather than modifying the compiler source code, we eliminate ``version hell'' and ensure that applications remain portable and standard-compliant.
Note that applications explicitly written for bare-metal (e.g., Rust \texttt{no\_std}) do not need ZeroOS or any ZeroOS-specific build config.

\subsection{Bootloader}

The bootloader handles three main responsibilities: (1) initialize platform components (e.g., CPU, memory, devices), (2) load and initialize the system (kernel, libc), then (3) transfer control to the application.
Any zkVM platform can easily support ZeroOS by implementing \texttt{\_\_platform\_bootstrap} for their specific platform (Figure~\ref{fig:zero-os-crates}).
Many zkVMs implement custom inlines and syscall instruction handling, which makes it impractical to include generic or default implementations of \texttt{\_\_platform\_bootstrap} in ZeroOS.

However, this design constraint isolates all platform-specific logic to this single function; all other components of ZeroOS and the ZeroOS bootloader are platform-agnostic. This minimal integration surface area allows the ecosystem to consolidate development and security auditing resources on the shared, generic ZeroOS kernel, rather than fracturing effort across completely custom runtimes.

\subsubsection{I/O}
\label{sec:design-devices}

ZeroOS supports peripherals, such as block devices and human interface devices using memory-mapped I/O (MMIO). On top of this low-level I/O layer, it is fairly straightforward to write device drivers for (logical) devices exposed by the zkVM platform.

For brevity, we only provide an overview of how MMIO is implemented in ZeroOS--port-mapped I/O is very similar, just using port-mapped I/O instructions rather than manipulating memory.

ZeroOS supports MMIO through the interface exposed by the zkVM platform - on a real hardware platform this would be a combination of the MMU and I/O device drivers, but on a zkVM platform this extends to more ad-hoc logical interfaces (Host-Target Interface (HTIF)).
Physical devices typically also require hardware interrupts (device to host) to handle asynchrony between the CPU and the device controller, but this is not strictly necessary in zkVM.
Optionally, syscalls such as \texttt{ioctl} are not part of the initial release of ZeroOS, but can be added as a module facilitate device management by the application.

The platform-specific \texttt{\_\_platform\_bootstrap} function bootstraps all devices in the zkVM platform; this function is necessarily different for each zkVM, and is the only zkVM-specific code that each zkVM project must maintain to support ZeroOS.
Once the devices are bootstrapped (control registers and/or buffers mapped into memory), they can be manipulated by device drivers or syscall handlers through the memory subsystem.

Note that for more complex devices, the zkVM must implement (deterministic) hardware interrupt handling between the device and the CPU.
Both the device and CPU are logical components of the zkVM platform, so while this would be complicated to implement it is not impossible in principle.

\subsection{Kernel}

\begin{figure}
    \centering
    \includegraphics[width=1.0\linewidth]{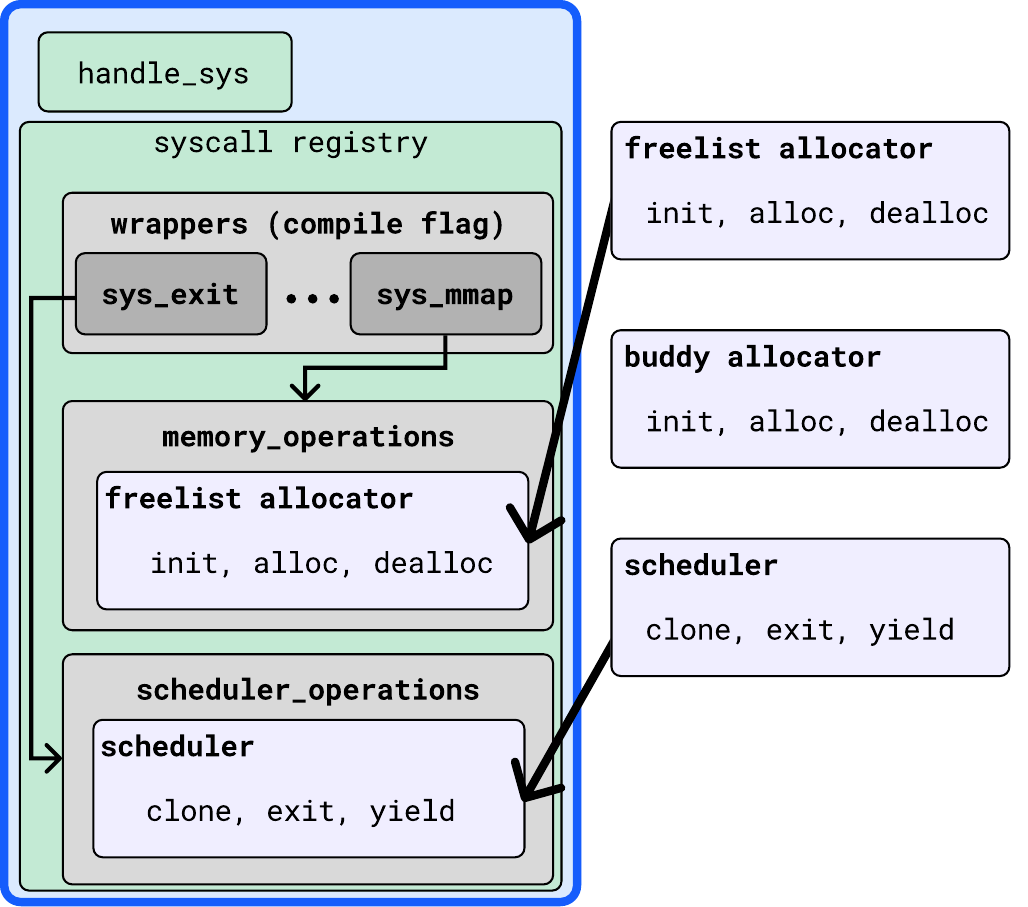}
    \caption{The ZeroOS kernel includes the trap handler (\texttt{handle\_sys}), wrapper interfaces for each syscall, and a set of imported syscall ops. The syscall ops primitives concretely implement all of the necessary low-level operations invoked by the abstract wrappers.}
    \label{fig:kernel-architecture}
\end{figure}

The ZeroOS kernel design is illustrated in Figure~\ref{fig:kernel-architecture}: it includes the trap handler (\texttt{handle\_sys}), abstract implementations of each syscall (\texttt{wrappers}) that call into a set of primitives stored in the operation registries (\texttt{memory\_operations} and \texttt{scheduler\_operations}).
The wrappers can be enabled and disabled via compile flags, so vApp developers can minimize the bytecode size of their ZeroOS unikernels.

Syscall wrappers define the high level logic of each syscall, calling into the operations registries to use low level primitives.
One example of a wrapper is \texttt{sys\_mmap}, which will call \texttt{memory\_operations.alloc}; the vApp developer can freely change the imported kernel memory allocator implementation (e.g., freelist vs buddy) to match their precise needs.
Wrappers are configured at compile time, and the operation registries are populated by the bootloader during system initialization.

Different applications may require different feature sets; for example, an application which does not use I/O may choose to exclude I/O-related syscalls to minimize compiled code size.
Compared to monolithic kernels, this modular design has superior security and efficiency in the context of vApps: it allows developers to ``unplug'' unused modules to strictly minimize the trusted computing base and execution trace length, achieving the efficiency of bare-metal optimization with the compatibility of a full OS.

Implementing features at the kernel level as opposed to the standard library level significantly reduces the chances of breaking changes.
For example, even in 2001 when Linux added 64-bit support to \texttt{mmap}, a 64-bit \texttt{mmap2} was added on top of the existing 32-bit \texttt{mmap} support with no breaking change to the ABI~\cite{mmap2}; this shows that kernel ABIs remain stable for decades, making this layer of the stack practical and desirable for integration.
This stability drives \emph{ecosystem consolidation}: instead of fragmenting security resources across version-specific toolchain patches, the industry can consolidate auditing efforts on a universal kernel where one fix immediately benefits all applications and zkVMs.

\begin{figure}[t]
    \centering
    \includegraphics[width=1.0\linewidth]{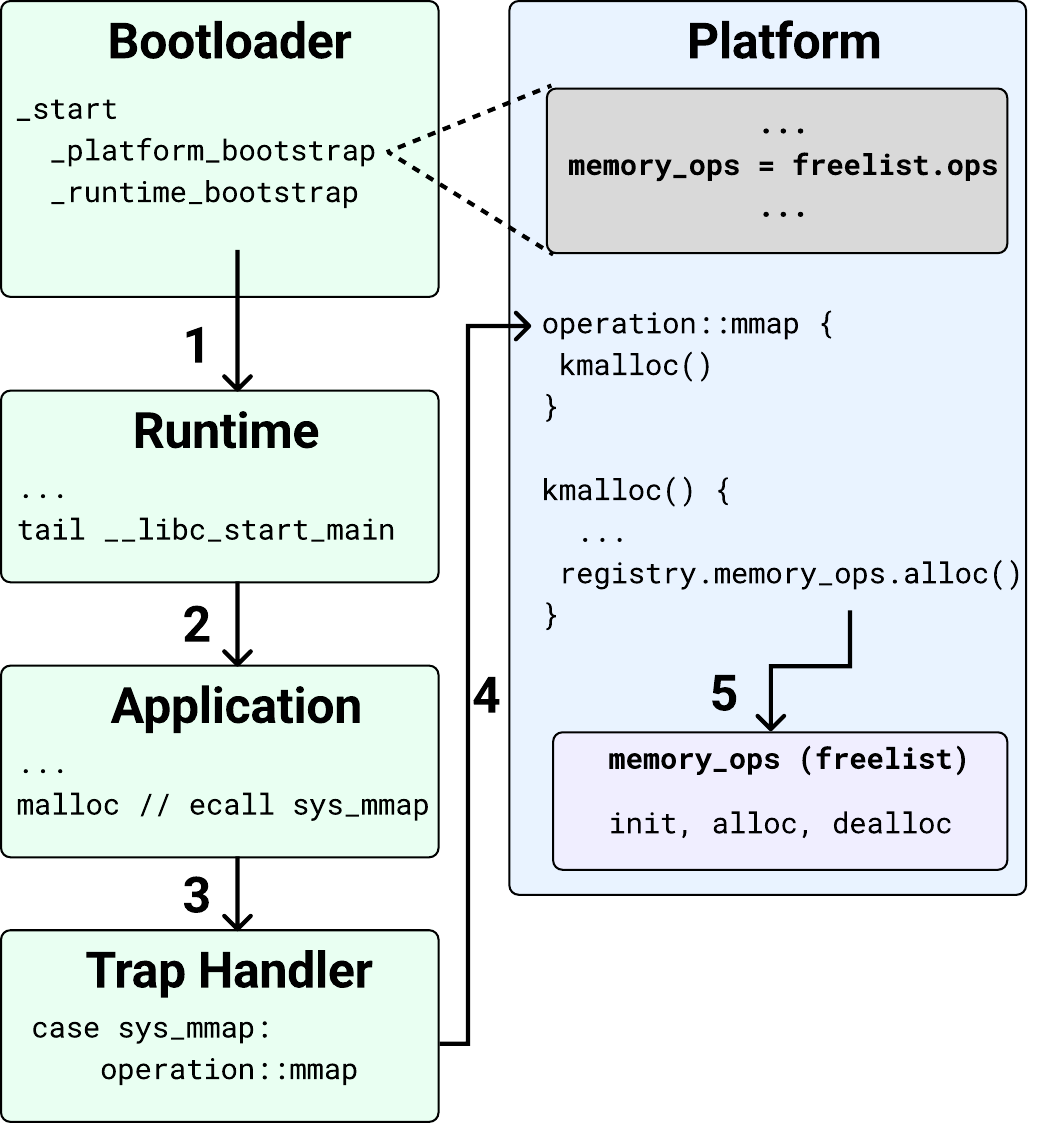}
    \caption{End-to-end execution flow to bootstrap a guest unikernel and call \texttt{sys\_mmap}. First, the bootloader bootstraps the platform, populating the platform \texttt{memory\_ops} with the freelist implementation. In call (1), the bootloader bootstraps and invokes the libc runtime (\texttt{\_\_libc\_start\_main}), which performs some bookkeeping before handing execution to the application in call (2). The application executes a syscall instruction (\texttt{ecall}) which traps into the ZeroOS trap handler in call (3). The trap handler calls the platform wrappers in call (4), which eventually calls into the \texttt{memory\_ops} alloc function in call (5). }
    \label{fig:e2e-mmap}
\end{figure}

\section{Implementation}
\label{sec:implementation}
Current vApps are primarily limited by prover cost, which is a direct function of the trace length.
This has led to the majority of zkVM and vApp developers simplifying their unikernels to minimize trace length.
As such, our initial release of ZeroOS is built with this goal in mind--ZeroOS supports \emph{single-process}, unified memory space unikernels with no simultaneous parallelism.
This initial implementation is highly modular, and all modules can easily be swapped out for more feature-rich modules as they are contributed to the codebase.
In Section~\ref{sec:discussion}, we discuss how additional features such as parallelism, virtual memory, and virtual filesystem (VFS) could be implemented in ZeroOS for future zkVMs.

To strike a good balance between minimizing the performance footprint of ZeroOS while still maintaining a familiar and full-featured interface for developers, we chose to target \texttt{musl}~\cite{musl} (unmodified) by implementing the Linux ABI/syscall interface in ZeroOS.
The implementation is highly modular and extensible, and supports OS + runtime composition (e.g., \texttt{linux-musl}, \texttt{linux-gnu}, \texttt{bsd}), and serves as an open platform for the community to contribute additional OS ABIs and runtimes.

The end-to-end flow from boot to allocating memory in \texttt{sys\_mmap} is outlined in Figure~\ref{fig:e2e-mmap}.
In this section, we go into detail about the implementation of each step.

\subsection{Bootloader}
The bootloader implements the following flow in ZeroOS: \texttt{\_start} $\rightarrow$ \texttt{\_\_platform\_bootstrap} $\rightarrow$ \texttt{\_\_runtime\_bootstrap} $\rightarrow$ \texttt{\_\_libc\_musl\_main} $\rightarrow$ application \texttt{main}.
The full boot sequence is included in Appendix~\ref{sec:appendix-boot-sequence}.

\vspace{1em}
\textbf{Step 0}: Hardware Reset - the zkVM is assumed to be in an untainted state at initialization

\vspace{1em}
\textbf{Step 1}: Kernel \texttt{\_start} sets up the global pointer and stack, then jumps to \texttt{\_\_bootstrap} (implemented in Rust).
This step is composed of two subcalls, first to \texttt{\_\_platform\_bootstrap} which
initializes any I/O devices and syscall shim (in RISC-V, by writing the address of the trap handler into the mtvec register).

Note that the syscall shim initialization is specific to the ISA of the zkVM, which often diverges from the RISC-V specification. For example, a RISC-V zkVM may use a fixed address for the syscall trap handler rather than storing the address in \texttt{mtvec}; in this case, \texttt{\_\_platform\_bootstrap} may be mostly empty. The source code corresponding to step 1 is included in Appendix~\ref{appendix-_start}.

\vspace{1em}
\textbf{Step 2}: The second subcall of \texttt{\_\_bootstrap} is \texttt{\_\_runtime\_bootstrap} (Appendix~\ref{appendix-runtime-bootstrap}), which builds the \texttt{musl} stack (Appendix~\ref{appendix-build-musl-stack}), runs pre-init/constructor hooks to prepare the stack for the application, then calls into \texttt{\_\_libc\_start\_main}.

\vspace{1em}
\textbf{Step 3}: \texttt{musl \_\_libc\_start\_main} invokes the application \texttt{main} and handles termination/cleanup.

\vspace{1em}
\textbf{Step 4}: Application \texttt{main} runs the developer's high level application logic.

\subsection{Kernel}
\subsubsection{Memory Management}
\paragraph{Memory model} The initial release of ZeroOS implements a single unified address space with no virtual memory or paging.
This unified address space is shared between the ZeroOS kernel and a single application (i.e., no process isolation).

\paragraph{Memory management model} ZeroOS uses a standard two-layer memory management hierarchy, with a \emph{userspace} allocator that handles guest memory allocations and a \emph{kernel} allocator that helps the userspace allocator expand or shrink its memory pools.

The userspace allocator is implemented in libc and the Rust standard library. Libc includes \texttt{malloc}, \texttt{free}, and \texttt{realloc}, and the Rust standard library implements \texttt{\_\_rust\_alloc} and \texttt{\_\_rust\_dealloc}.

The kernel allocator implements \texttt{mmap} and \texttt{munmap} on pages (e.g., 4\,KiB) of global physical memory (i.e., the region of memory described by \texttt{\_\_heap\_start} and \texttt{\_\_heap\_end}).

The top section of Table~\ref{tab:syscalls} describes the various memory-related syscalls used by libc for memory management, along with a discussion of which syscalls are required vs optional.

\renewcommand{\arraystretch}{1.2}
\begin{table*}
    \centering
    \begin{tabular}{@{}lp{0.55\textwidth}c@{}}
        \toprule
        \textbf{Syscall} & \textbf{Description} & \textbf{Implementation} \\
        \midrule
        \multicolumn{3}{@{}c@{}}{\textbf{Memory management}} \\
        \midrule
        \texttt{sys\_mmap} & Kernel heap allocator locates a free range and returns the base address to the caller. & Yes \\
        \texttt{sys\_munmap} & Returns a previously \texttt{mmap}'ed memory region to the kernel. & Yes \\
        \texttt{sys\_brk} & Adjusts the program break of the caller; superseded by \texttt{mmap}. & Stub (\texttt{-ENOMEM}) \\
        \texttt{sys\_mremap} & Resizes an existing \texttt{mmap}'ed region. & Stub (\texttt{-ENOSYS}) \\
        \texttt{sys\_mprotect} & Changes \texttt{mmap} region protection flags; a no-op in our single-tenant unikernel. & Stub (no-op) \\
        \midrule \multicolumn{3}{@{}c@{}}{\textbf{Thread management}} \\
        \midrule
        \texttt{sys\_clone} & Fundamental primitive for creating new threads; allocates a TCB and sets up the thread for execution. & Yes \\
        \texttt{sys\_futex} (\texttt{FUTEX\_WAIT}) & Checks the futex word, records the thread as blocked on the futex, and yields to the next thread. & Yes \\
        \texttt{sys\_futex} (\texttt{FUTEX\_WAKE}) & Moves all threads blocked on a given futex to the ready queue and returns the number woken. & Yes \\
        \texttt{sys\_sched\_yield} & Allows a thread to yield its time slice, pushing the caller onto the end of the ready queue. & Yes \\
        \texttt{sys\_exit} & Terminates the calling thread, deallocates its TCB, and removes it from the ready queue. & Yes \\
        \texttt{sys\_exit\_group} & Shuts down the process group; in a ZeroOS unikernel this terminates the whole system. & Yes \\
        \texttt{sys\_set\_tid\_address} & Sets the Thread ID (TID) pointer for the calling thread. & Stub (no-op) \\
        \texttt{sys\_rt\_sigaction} & Change the handler for a given signal. ZeroOS does not support signals, so this syscall is stubbed. & Stub (no-op) \\
        \texttt{sys\_rt\_sigprocmask} & Manipulate the signal mask for a given process. ZeroOS does not support signals, so this syscall is stubbed. & Stub (no-op) \\
        \midrule \multicolumn{3}{@{}c@{}}{\textbf{I/O}} \\
        \midrule
        \texttt{sys\_write} & Writes to a file descriptor. In ZeroOS, the implementation is dependent on the devices initialized during bootstrapping. & stdout / stderr \\
        \texttt{sys\_fstat} & Return information about a given file descriptor. ZeroOS supports fstat for all platform devices. & stdout / stderr \\
        \texttt{sys\_writev} & Write multiple buffers. & stdout / stderr \\
        \texttt{sys\_prlimit64} & Get or set resource limits. In our single-tenant system, resource limits are superfluous. & Stub (no-op) \\
        \texttt{sys\_getrandom} & Return a random value. Real entropy requires platform support, ZeroOS currently returns a non-random value. & Return dummy value \\
        \texttt{sys\_clock\_gettime} & Return current system time. Real time requires platform support, ZeroOS defaults to returning the zero time. & Return zero time \\
        \texttt{sys\_ioctl} & Manage I/O devices. ZeroOS does not currently support specialized I/O devices, only memory-mapped stdout and stderr via platform-defined interfaces. & Stub (\texttt{-ENOTTY}) \\
        \bottomrule
    \end{tabular}
    \caption{ZeroOS syscall interface.}
    \label{tab:syscalls}
\end{table*}
\renewcommand{\arraystretch}{1.0}

When designing the ZeroOS kernel heap allocator, we evaluated several designs with respect to the key concerns of zkVM: auditability, determinism, predictability, feature-sufficiency, and runtime complexity.
In particular, \texttt{sys\_munmap} support is required for compatibility with most modern implementations of \texttt{libc}, as allocations larger than \texttt{DEFAULT\_MMAP\_THRESHOLD\_MIN} (128\,KiB in \texttt{musl}) bypass the user-space allocator entirely and use \texttt{sys\_mmap} and \texttt{sys\_munmap} directly.

\begin{table*}
    \centering
    \begin{tabular}{@{}llllll@{}}
        \toprule
        \textbf{Allocator} & \textbf{Deallocation} & \textbf{Simplicity} & \textbf{Fragmentation} & \textbf{Small objects} & \textbf{Runtime complexity} \\
        \midrule
        Bump allocator & No & Simplest & N/A & Poor & Best \\
        Slab allocator & Yes & Excellent & Minimal & Excellent & Excellent \\
        \texttt{kmalloc}-style & Yes & Complex & Minimal & Excellent & Excellent \\
        (SLAB/SLUB)  & & & & & \\
        Coalescing free list & Yes & Excellent & Vulnerable to & Moderate & Good \\
        & & & adversarial workloads & & \\
        Buddy allocator & Yes & Complex & Reasonable & Good & Good \\
        \bottomrule
    \end{tabular}
    \caption{\textbf{Kernel allocator design space.} Simpler allocators are easier to audit but may suffer from fragmentation or limited feature support; ZeroOS uses a coalescing free list as a balanced choice.}
    \label{tab:allocators}
\end{table*}
We choose the coalescing free list (e.g., \texttt{linked\_list\_allocator::LockedHeap}) for the first ZeroOS kernel allocator, as it strikes a good balance with our key concerns (Table~\ref{tab:allocators}.
Alternatively, the bump allocator is simpler with no fragmentation, but does not support reclamation.
We believe a buddy allocator is a viable alternative if future workloads demand stricter bounds on fragmentation, but the additional code and runtime complexity makes this a long-term target.

Note that applications will always be more efficient if they manage memory in userspace, so the expectation is that skilled developers will minimize calls to \texttt{sys\_mmap} to reduce trace length and consequently prover cost.

\subsubsection{Thread management}

\begin{figure}
    \centering
    \includegraphics[width=1.0\linewidth]{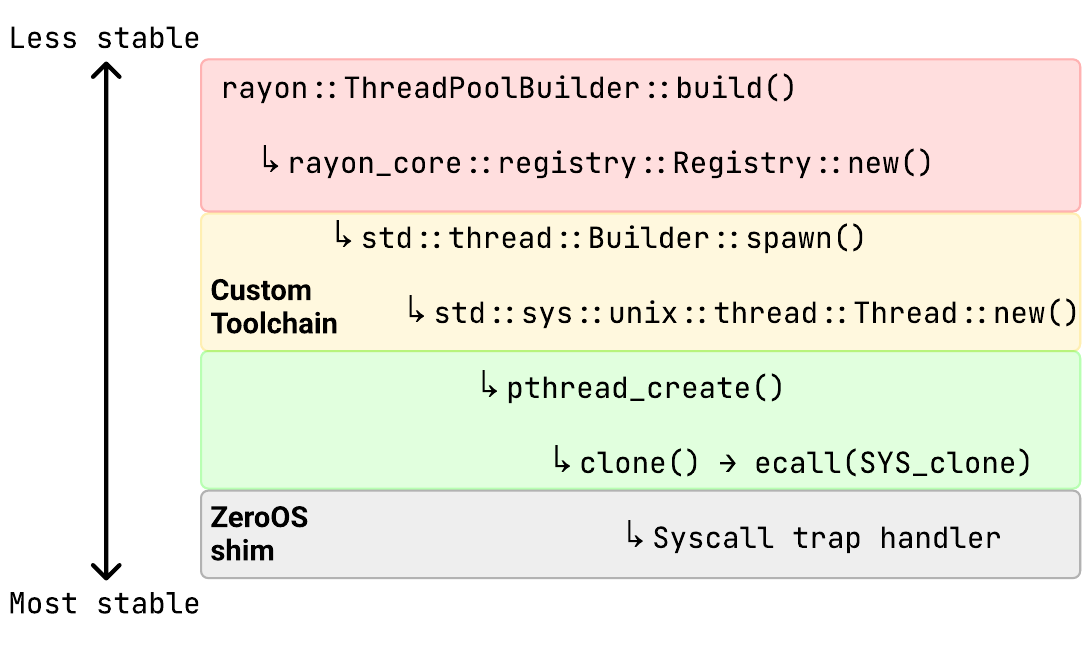}
    \caption{ZeroOS is at the bottom of the callstack, right above the bare-metal, ensuring the highest stability to minimize ``version hell''. The current status quo of using custom forked Rust \texttt{std} incurs higher risk of breaking changes between versions.}
    \label{fig:stable-dependencies}
\end{figure}

Modern threading libraries (e.g., rayon~\cite{rayon}) are typically built on threading primitives exposed by \texttt{libc} which in turn build on syscalls such as \texttt{sys\_futex}.
In ZeroOS, we fully implement five main threading-related syscalls intended to enable basic multithreading: \texttt{sys\_clone}, \texttt{sys\_futex}, \texttt{sys\_sched\_yield}, \texttt{sys\_exit}, and \texttt{sys\_exit\_group}.
These five syscalls provide the bare minimum functionality to implement a non-preemptive multithreading environment for a single-application unikernel.
Additionally, we stub three thread management syscalls (\texttt{sys\_set\_tid\_address}, \texttt{sys\_rt\_sigaction}, and \texttt{sys\_rt\_sigprocmask}) to allow threading libraries to compile properly.
Other thread management syscalls such as \texttt{gettid}, \texttt{sched\_\{set,get\}*}, and multitasking syscalls such as \texttt{fork} and \texttt{exec} may be implemented in future versions of ZeroOS as the need arises.
Certain thread management syscalls are not implemented such as \texttt{sys\_tgkill} as ZeroOS does not use Unix signals for control flow.

The middle section of Table~\ref{tab:syscalls} outlines the threading-related syscalls in the ZeroOS design.

Threading and multitasking in a traditional kernel is often quite complex and heavyweight.
For example, the thread control block (TCB) (i.e., \texttt{task\_struct}) tracks memory mappings, file descriptors, user/group identities, signal handlers, scheduling policies, and more; the definition of \texttt{task\_struct} in the Linux kernel contains over 300 fields~\cite{linux-task-struct}, most of which are unnecessary for zkVM guests.
ZeroOS assumes a single process, a single address space, an no application-level parallelism in its current implementation.
All threads share the same memory and file system context, and there is no notion of per-process credentials or per-process address spaces.
This allows us to use a radically simplified TCB that contains only what is needed for scheduling and context switching, shown in Table~\ref{tab:tcb}.

\begin{table}
    \centering
    \begin{tabular}{@{}lp{0.6\linewidth}@{}}
        \toprule
        \textbf{Field name} & \textbf{Description} \\
        \midrule
        \texttt{saved\_regs} &
        Thread's program counter, stack pointer, and general-purpose registers. \\
        \texttt{thread\_state} &
        Thread's state enum: \texttt{Ready}, \texttt{Blocked}, or \texttt{Exited}. \\
        \texttt{tid} &
        Thread ID (returned by \texttt{sys\_clone} and \texttt{gettid}). \\
        \bottomrule
    \end{tabular}
    \caption{\textbf{ZeroOS TCB fields.} Minimal per-thread state needed for scheduling and context switching.}
    \label{tab:tcb}
\end{table}

These five syscalls and minimal TCB enable ZeroOS to support the \texttt{std::thread} API and \texttt{rayon}'s work-stealing scheduler without compromising on our primary concerns of auditability, determinism, predictability, feature-sufficiency, and runtime complexity.
An example of the resulting callstack of an invocation to \texttt{rayon::ThreadPoolBuilder::build()} is shown in Figure~\ref{fig:stable-dependencies}.
Custom toolchains often inject their logic in the shallower, less stable interfaces within the callstack, whereas ZeroOS uses unmodified toolchains and code until the bottom of the callstack.

\subsection{I/O}
\label{sec:implementation-io}
ZeroOS supports rudimentary I/O in the form of stdout and stderr via memory-mapped I/O.
Both stdout and stderr and initialized during platform bootstrapping, and are permanently mapped to file descriptors 1 and 2 respectively.
The bottom section of Table~\ref{tab:syscalls} lists the I/O-related syscall interface of ZeroOS; \texttt{sys\_write}, \texttt{sys\_fstat} and \texttt{sys\_writev} are implemented to handle stdout and stderr.
Additional syscalls such as \texttt{sys\_prlimit64}, \texttt{sys\_getrandom}, \texttt{sys\_clock\_gettime}, and \texttt{sys\_ioctl} are stubbed to resolve compilation issues, but are not functionally necessary for current vApps.

ZeroOS interfaces with stdout and stderr via the \emph{Host-Target Interface} (HTIF) defined by the platform and initialized during platform bootstrapping.
In the current implementation of ZeroOS, we support an HTIF that exposes stdout and stderr as reserved regions of memory where debug output from the program is written.
The HTIF defines the interface by which ZeroOS target interacts with the zkVM host, and the responsibility of actually passing this information to the end user (i.e. developer) is outside the scope of ZeroOS.

Note that I/O in the current version of ZeroOS (specifically the HTIF we implement for the RISC-V Spike~\cite{spike} platform) is intended to be used strictly as a debugging aid, as it is not strictly deterministic and therefore may complicate guest security analysis.

\subsection{Build Configuration}
The build configuration is responsible for linking the application, libraries, and ZeroOS kernel into a single executable using a specialized build pipeline.

\begin{enumerate}
    \item Toolchain preparation: Build static \texttt{libc.a} using any \texttt{musl}-based RISC-V toolchain
    \item Target specification: Specify a custom RISC-V target json that links the guest application against \texttt{libstd} and \texttt{musl} using the Linux ABI.
    \item Static linking: The linker resolves symbols so that the Rust standard library calls into \texttt{musl}, and \texttt{musl} syscall wrappers properly trap into the ZeroOS kernel syscall shim.
\end{enumerate}

This process outputs a unikernel, which is a self-contained ELF that can run on bare-metal in zkVM while retaining the functionality of an application running on Linux.

\section{Discussion and Future Work}
\label{sec:discussion}
The ZeroOS implementation presented in this paper establishes the core primitives for a verifiable library OS. While the initial release targets single-process unikernels, the modular architecture effectively crowdsoruces the operating system's evolution.

Prior work by Tsai et al.~\cite{10.1145/2901318.2901341} has suggested that supporting 100\% of applications in a typical debian distribution requires implementing 272 system calls, but only 81 for the top 10\% most popular applications (and 40 for the top 1\%).
Additionally, prior work by Unikraft has shown that as much as 60\% of syscalls can be faked/stubbed without affecting traditional workloads~\cite{unikraft-compatibility}.


In this section, we discuss how ZeroOS facilitates the community-driven implementation of these additional features. Once contributed, these modules become instantly available to every platfrom in the ecosystem.

\subsection{Virtual Memory and Paging}
Supporting virtual memory in ZeroOS is feasible within the existing shim architecture, although it significantly increases the code and runtime complexity of the kernel.
Many syscalls must be implemented or modified, including but not limited to \texttt{mmap}, \texttt{munmap}, \texttt{mlock}, \texttt{mprotect}, \texttt{mincore}, etc.
In particular, many functions such as \texttt{mmap} must maintain page tables in a way that is compatible with the ISA-defined MMU, and other small maintenance tasks such as TLB invalidation.

Virtual memory can be implemented in ZeroOS much in the same way it would be implemented in any other operating system, provided the zkVM exposes the necessary platform-level primitives such as the MMU.

\subsection{Multitasking and IPC}
Multitasking follows the established modular pattern. The ZeroOS design already establishes the foundation with non-preemptive multithreading primitives based on \texttt{sys\_clone} and \texttt{sys\_futex}; full multi\emph{tasking} would extend this foundation using \texttt{sys\_fork}, \texttt{sys\_wait4}, and \texttt{sys\_execve}.

Inter-Process Communication (IPC) can be implemented via message queues (\texttt{sys\_mq\_*}, etc.), shared memory (\texttt{sys\_shm\_*}, etc.), pipes (\texttt{sys\_pipe}, etc.), or network.
We discuss how the VFS can be implemented in Section~\ref{sec:discussion-vfs}, which enables IPC via pipes and network interfaces.
Message queues and shared memory can be implemented efficiently over a memory buffer that is either managed by the message queue syscall interface or directly exposed as shared memory, preserving the kernel's low footprint.

\subsection{Signaling, time, and randomness}
Signaling is currently not supported in ZeroOS (e.g., for preemptive scheduling) because current zkVMs do not implement deterministic hardware signals; this is a difficult problem to solve as signals are intended to be nondeterministic, so injecting determinism would constitute a divergence from the ISA.
However, assuming a zkVM supports deterministic signaling primitives, ZeroOS can build upon those deterministic signaling primitives and invariants to implement all signal-related syscalls.

Time and randomness face similar constraints. While the syscalls are trivially implementable given deterministic interfaces, the overriding concern is protecting the vApp from a malicious prover. All inputs must be fully constrained to prevent the prover from manipulating the execution trace. As a simplified example, ZeroOS can enforce that time and random numbers are deterministically derived from public inputs, thereby removing any degree of freedom for malicious manipulation.

\subsection{Virtual Filesystem and Network}
\label{sec:discussion-vfs}
The virtual filesystem (VFS) interface can be implemented by simply maintaining an \texttt{fdtable} that corresponds to memory-mapped I/O devices or an in-memory backing storage (e.g., ramdisk).

Given I/O and VFS support, the sockets interface for network support is quite straightforward.
Network devices would be initialized as part of the platform, managed via MMIO or PMIO, and exposed via VFS. Cruicially, adhering to the Pareto Principle, these heavy subsystems remain strictly optional within the ZeroOS architecture. A vApp that requires networking can link the module, while a pure computation vApp can exclude it entirely, ensuring that the "pay-for-what-you-use" efficiency is preserved even as the OS grows.

\section{Conclusion}

ZeroOS is a necessary maturation of the zkVM software stack.
We eliminate the ``version hell` of custom toolchains and enable universal language and platform compatibility by moving system-level logic from fragile language runtimes into a language-agnostic Library OS.
ZeroOS enables compilation of existing applications into vApps using off-the-shelf toolchains with a simple configuration change, and any zkVM can join the ZeroOS ecosystem by implementing a minimal platform bootstrapper.

Adopting the modular unikernel model delivers a Pareto improvement over the status quo.
It allows developers to construct guests that possess the high-level compatibility of a full OS while retaining the ability to ``unplug'' modules to strictly minimize the trusted computing base.

Our current implementation prioritizes single-process unikernels with non-preemptive multithreading to minimize prover cost, and the ZeroOS architecture lays a robust foundation for future expansion.
In our future work, we plan to explore complex features such as simultaneous multitasking, virtual memory, and VFS support through ZeroOS modules to support a wider variety of workloads without unnecessarily expanding the trusted computing base for simpler applications.
Additionally, we hope that as zkVM platforms support a richer set of I/O-related primitives, we can shift from limited HTIF-defined I/O interfaces to more general support for MMU and device drivers in ZeroOS.

Finally, ZeroOS achieves higher practical security by consolidating ecosystem resources.
ZeroOS provides a single platform to consolidate the fragmented landscape of ad-hoc patches with a unified operating system, and shifts the security model from a function of individual team resources to a function of the collective ecosystem.
This unification simplifies development, hardens security, and accelerates the adoption of verifiable application.

\bibliographystyle{acm}
\bibliography{bib}

\onecolumn
\appendix

\section{ZeroOS boot sequence}
\label{sec:appendix-boot-sequence}
\begin{lstlisting}[style=CallStack]
Hardware reset
    CALLINTO `_start` (RISC-V, .text.boot)
        CALLINTO `__bootstrap`
            CALLINTO `__platform_bootstrap`
                CALLINTO register trap-handler
            CALLINTO `__runtime_bootstrap`
                CALLINTO `build_musl_stack(buffer_top, ehdr_start, PROGRAM_NAME)`
                    CALLINTO construct stack: argc / argv / envp / auxv
                CALLINTO `__libc_start_main(main, argc, argv, _init, _fini, NULL)`
                    CALLINTO `__init_libc(envp, argv[0])`
                        CALLINTO parse `auxv`, set `libc.page_size`, `__hwcap`, etc.
                    CALLINTO `__init_tls(aux)`
                    CALLINTO `__init_ssp((void*)aux[AT_RANDOM])`
                    CALLINTO `libc_start_main_stage2(main, argc, argv)`
                        CALLINTO `__libc_start_init()`
                        CALLINTO `__main_entry(argc, argv)`
                            CALLINTO user main()
\end{lstlisting}


\section{\texttt{\_start}}
\label{appendix-_start}

\begin{minted}{rust}
/// Boot entry point - shared across all libc implementations.
///
/// Sets up GP and SP, then tail-calls __bootstrap for the rest of initialization.
#[unsafe(naked)]
#[link_section = ".text.boot"]
#[no_mangle]
pub unsafe extern "C" fn _start() -> ! {
    naked_asm!(
        // Initialize global pointer first (RISC-V ABI requirement)
        ".weak __global_pointer$",
        ".hidden __global_pointer$",
        ".option push",
        ".option norelax",
        "   lla     gp, __global_pointer$",
        ".option pop",

        // Initialize stack pointer
        ".weak __stack_top",
        ".hidden __stack_top",
        "   lla     sp, __stack_top",
        "   andi    sp, sp, -16",

        // Tail call to bootstrap coordinator
        "   tail    {bootstrap}",

        bootstrap = sym __bootstrap,
    )
}

/// Bootstrap coordinator - orchestrates the initialization sequence.
/// 1. __platform_bootstrap - platform hardware setup
/// 2. __runtime_bootstrap - runtime environment setup (libc stack or no-std)
#[unsafe(naked)]
#[no_mangle]
pub unsafe extern "C" fn __bootstrap() -> ! {
    naked_asm!(
        // 1. Platform initialization (trap handler, etc.)
        "   call    {platform_bootstrap}",
        
        // 2. Runtime environment initialization (never returns)
        "   tail    {runtime_bootstrap}",

        // Safety: If main() returns, halt forever.
        // User should call exit() to terminate properly.
        // Using inline asm instead of `loop {}` or `spin_loop()` because:
        // - `loop {}` may be optimized away as undefined behavior
        // - `spin_loop()` generates `pause` only with Zihintpause extension,
        //   otherwise no instruction on RISC-V
        // - `j .` guarantees a single-instruction infinite loop
        "   j       .",
        
        platform_bootstrap = sym __platform_bootstrap,
        runtime_bootstrap = sym __runtime_bootstrap,
    )
}
\end{minted}

\section{\texttt{\_\_runtime\_bootstrap}}
\label{appendix-runtime-bootstrap}
\begin{minted}{rust}
#[unsafe(naked)]
#[no_mangle]
pub unsafe extern "C" fn __runtime_bootstrap() -> ! {
    naked_asm!(
        // Build musl stack in buffer (returns buffer_sp in a0)
        "   call    {build_impl}",
        
        // Snapshot current platform SP after helper returns
        "   mv      t4, sp",
        
        // Calculate buffer_top, size, and target_sp
        // t0 = buffer_top, t1 = buffer_bytes, t2 = size
        "   la      t0, {buffer}",
        "   li      t1, {buffer_bytes}", // t1 = buffer size in bytes
        "   add     t0, t0, t1",         // t0 = buffer_top
        "   sub     t2, t0, a0",         // t2 = size (buffer_top - buffer_sp)
        
        // Calculate target_sp (where to copy the stack)
        // t3 = target_sp (current_sp - size)
        "   sub     t3, t4, t2",         // t3 = target_sp
        "   mv      t5, t3",             // t5 = saved target_sp for final SP
        
        // Copy loop: copy from buffer to target location
        // t6 = src (buffer_sp), t3 = dst (current position), t2 = remaining bytes
        "   mv      t6, a0",             // t6 = src (buffer_sp)
        "1:",
        "   beqz    t2, 2f",             // if size == 0, done
        "   ld      t1, 0(t6)",          // load 8 bytes from buffer
        "   sd      t1, 0(t3)",          // store 8 bytes to target
        "   addi    t6, t6, 8",          // src += 8
        "   addi    t3, t3, 8",          // dst += 8
        "   addi    t2, t2, -8",         // size -= 8
        "   j       1b",                 // loop
        "2:",
        
        // Switch to the new musl stack (use saved target_sp)
        "   mv      sp, t5",             // sp = target_sp (start of copied stack)
        
        // Now SP points to musl stack with layout:
        //   SP+0: argc
        //   SP+8: argv[0]
        //   SP+16: argv[1]
        //   ...
        
        // Load argc and argv from musl stack
        "   lw      a1, 0(sp)",          // a1 = argc
        "   addi    a2, sp, 8",          // a2 = argv
        
        // Load function pointers for __libc_start_main
        "   la      a0, {main}",         // a0 = main function
        "   la      a3, {init}",         // a3 = _init
        "   la      a4, {fini}",         // a4 = _fini
        "   li      a5, 0",              // a5 = NULL (ldso_dummy)
        
        // Tail call __libc_start_main (never returns)
        "   tail    {libc_start_main}",
        
        build_impl = sym build_musl_in_buffer,
        buffer = sym MUSL_BUILD_BUFFER,
        buffer_bytes = const MUSL_BUFFER_BYTES,
        main = sym __main_entry,
        init = sym _init,
        fini = sym _fini,
        libc_start_main = sym __libc_start_main,
    )
}
\end{minted}

\section{\texttt{build\_musl\_stack}}
\label{appendix-build-musl-stack}
\texttt{build\_musl\_stack} pushes the required \texttt{auxv} entries onto the stack.
Note that ZeroOS must set \texttt{AT\_PHNUM} (program header count) to 0 to ensure \texttt{musl} skips reading the ELF headers (which may not be accessible) and instead use the provided defaults.

\begin{minted}{rust}
pub unsafe fn build_musl_stack(
    _initial_sp: usize,
    _ehdr_start: usize, // Ignored - ELF header not accessible in bare-metal
    program_name: &'static [u8],
) -> usize {
    let mut ds = DownwardStack::<usize>::new(_initial_sp);

    // Zero-auxv approach: Tell musl that program headers are not available.
    // Musl will skip reading PT_TLS from the ELF header when AT_PHNUM=0.
    let at_phdr = 0; // No program headers available
    let at_phent = 0; // No program header entry size
    let at_phnum = 0; // No program header count

    // Prepare auxiliary vector entries
    let auxv_entries = [
        (AT_PHDR, at_phdr),
        (AT_PHENT, at_phent),
        (AT_PHNUM, at_phnum),
        (AT_PAGESZ, 4096),
        (AT_CLKTCK, 100),
        (AT_HWCAP, 0),
        (AT_UID, 0),
        (AT_EUID, 0),
        (AT_GID, 0),
        (AT_EGID, 0),
        (AT_SECURE, 0),
        (AT_RANDOM, 0), // Will be replaced with actual pointer below
        (AT_NULL, 0),
    ];

    unsafe {
        // Generate 16 bytes for AT_RANDOM (Linux kernel standard)
        // Musl's __init_ssp uses first sizeof(uintptr_t) bytes for stack canary
        // Use multiple entropy sources: sp and a constant
        let entropy = [_initial_sp as u64, 0xdeadbeef_cafebabe_u64];
        let (random_low, random_high) = generate_random_bytes(&entropy);

        // Push 16 bytes in architecture-appropriate chunks
        //
        // Memory layout after pushes (stack grows downward, lower addresses at bottom):
        //
        // 32-bit architectures (4-byte usize):
        //   sp+12: random_high[63:32]  (4 bytes)
        //   sp+8:  random_high[31:0]   (4 bytes)
        //   sp+4:  random_low[63:32]   (4 bytes)
        //   sp+0:  random_low[31:0]    (4 bytes) ← at_random_ptr, musl reads 4 bytes
        //
        // 64-bit architectures (8-byte usize):
        //   sp+8:  random_high[63:0]   (8 bytes)
        //   sp+0:  random_low[63:0]    (8 bytes) ← at_random_ptr, musl reads 8 bytes
        //
        #[cfg(target_pointer_width = "32")]
        {
            ds.push((random_high >> 32) as usize); // High 32 bits of random_high
            ds.push(random_high as usize); // Low 32 bits of random_high
            ds.push((random_low >> 32) as usize); // High 32 bits of random_low
            ds.push(random_low as usize); // Low 32 bits of random_low (musl reads from here)
        }
        #[cfg(target_pointer_width = "64")]
        {
            ds.push(random_high as usize);
            ds.push(random_low as usize); // Musl reads from here
        }

        let at_random_ptr = ds.sp();

        // Push auxiliary vector (in reverse order since stack grows downward)
        for &(key, val) in auxv_entries.iter().rev() {
            let eff_val = if key == AT_RANDOM { at_random_ptr } else { val };
            ds.push(eff_val);
            ds.push(key as usize);
        }

        // Push envp (empty, NULL-terminated)
        ds.push(0);

        // Push argv (program name + NULL terminator)
        ds.push(0);
        ds.push(program_name.as_ptr() as usize);

        // Push argc
        ds.push(1);
    }

    ds.sp()
}
\end{minted}
\twocolumn

\end{document}